\documentclass[onecolumn,amsmath,amssymb,10pt]{revtex4}

\usepackage{graphicx}
\usepackage{dcolumn}
\usepackage{bm}
\usepackage{subfigure}
\usepackage{amsmath}
\usepackage{amsfonts}
\usepackage{amssymb}

\usepackage{subfigure}
\usepackage[nice]{nicefrac}
\usepackage[tight]{units}

\begin{document}

\title{Intensity Dependence of Laser-Assisted Attosecond Photoionization Spectra}

\author{M. Swoboda}
\email{marko.swoboda@fysik.lth.se}
\homepage{http://www.atto.fysik.lth.se}
\author{J. M. Dahlstr\"om}
\author{T. Ruchon}
\altaffiliation{now at: CEA-Saclay, DSM, Service des Photons, Atomes et Mol\'ecules, 91191 Gif sur Yvette, France}
\author{P. Johnsson}
\author{J. Mauritsson}
\author{A. L'Huillier}
\affiliation{Department of Physics, Lund University, P.O. Box 118, 22100 Lund, Sweden}
\author{K. J. Schafer}
\affiliation{Department of Physics and Astronomy, Louisiana State University, Baton Rouge, Louisiana 70803-4001, USA}

\date{\today}

\begin{abstract}
We study experimentally the influence of the intensity of the infrared (IR) probe field on attosecond pulse train (APT) phase measurements performed with the RABITT method (Reconstruction of Attosecond Beating by Interference in Two-Photon Transitions). We find that if a strong IR field is applied, the attosecond pulses will appear to have lower-than-actual chirp rates. We also observe the onset of the streaking regime in the breakdown of the weak-field RABITT conditions. We perform a Fourier-analysis of harmonic and sideband continuum states and show that the mutual phase relation of the harmonics can be extracted from higher Fourier components. 
\end{abstract}
\maketitle
\pagebreak
\section{Introduction}
Laser-assisted ionization processes provide an elegant tool to study dynamics and details of atomic and molecular systems \cite{CavalieriNature2007,DrescherNature2002,MauritssonPRL2008}. For time-dependent measurements, the assisting laser needs to be synchronous to the ionization event on a time scale shorter than that of the process to be resolved. A number of experiments in the past years have shown that the attosecond (as) time scale is routinely accessible and laser-assisted ionization processes have become the chief tool for experiments in the field of attosecond science \cite{CavalieriNature2007,DrescherNature2002,GoulielmakisScience2004,MauritssonPRL2008,JohnssonPRL2007,RemetterNP2006}.\par
The emission of a comb of high-order harmonics when an atomic medium is exposed to a driving intense laser field is well understood. The resulting attosecond pulse trains (APT) \cite{PaulScience2001,MairesseScience2003} provide a premier tool to controllably ionize atomic media \cite{JohnssonPRL2007,GuyetandJPhysB2008} and it is important to determine their characteristics, both amplitude and phase. This can be done by cross-correlation with an IR probing field under stable interferometric conditions. Using weak infrared (IR) fields, this is often done with the Reconstruction of Attosecond Beating by Interference of Two-Photon Transitions (RABITT) method \cite{PaulScience2001}, while the AC-Streak camera method \cite{ItataniPRL2002,KienbergerScience2002}, using stronger IR fields, has mainly been  applied to the characterization of isolated attosecond pulses. A special case 
is the full characterization method FROG-CRAB (Frequency-Resolved Optical Gating- Complete Reconstruction of Attosecond Bursts), which is based on an iterative deconvolution of a time-frequency spectrogram \cite{MairessePRA2005,SansoneScience2006}. In this paper, we study the transition from the weak to the strong field regime, and the effect of the probe field strength on the validity of the characterization. \par
In Section II, the experimental setup and performed experiments will be presented. We perform in Section III a conventional analysis of the data in the RABITT regime, studying specifically the probe intensity dependence of the phase measurement. A Fourier-series approach (Section IV) allows us to see the fingerprints of processes with more than one contributing IR photons and provides a more general description of our delay-dependent laser-assisted photoionization spectra.
\section{Experiment}
\subsection{Setup}
\begin{figure}
	\includegraphics[width=0.65\textwidth]{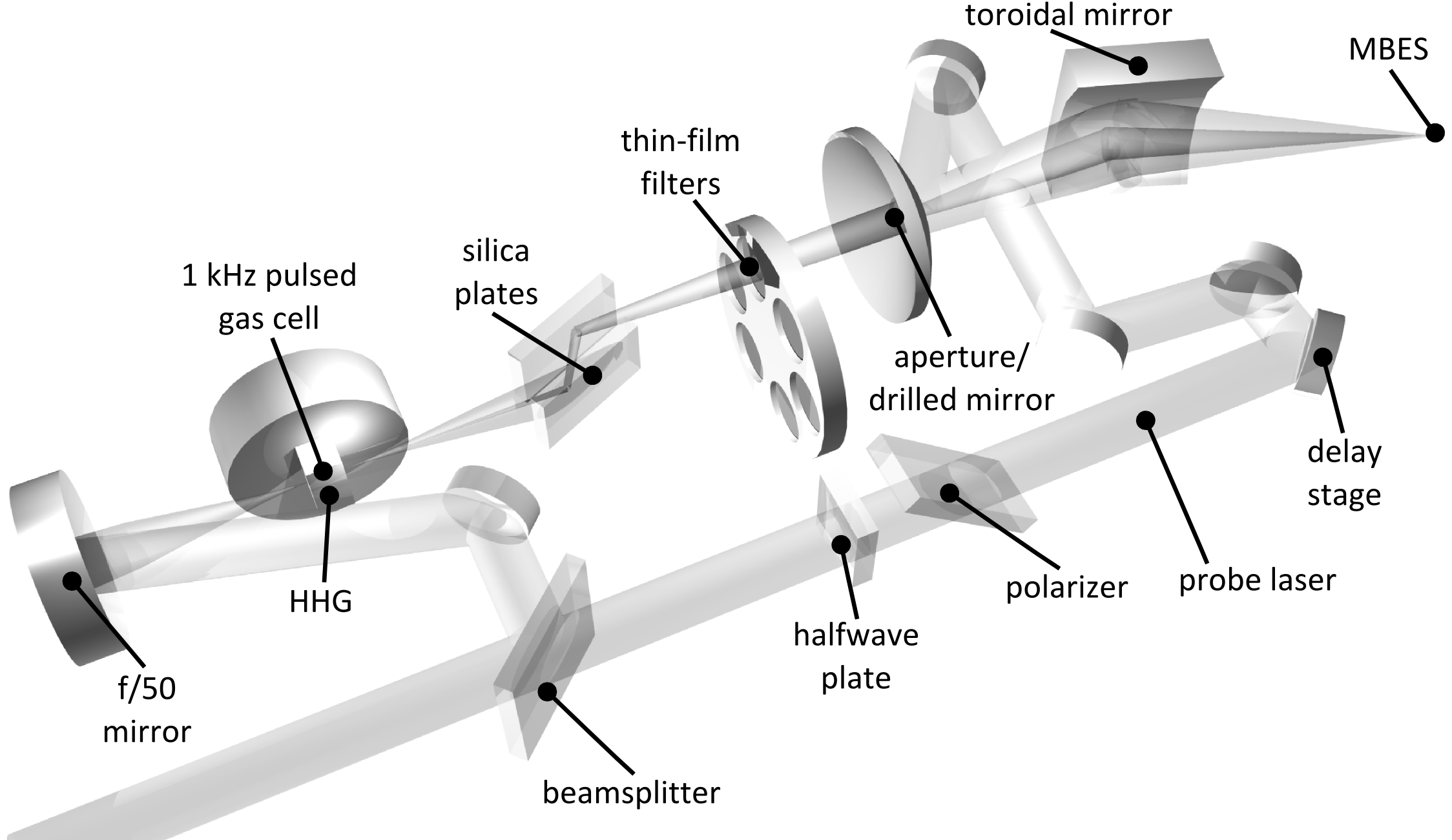}
	\caption{Setup for the experiment: a 2\,mJ, 30\,fs laser pulse is split into a probe and a pump arm. The larger part is used in high-order harmonic generation while a  fraction travels through a delay stage to serve as probe pulse. The generated harmonics are filtered spectrally and spatially by silica plates, a thin-film metallic filter and an aperture, before recombining with the probe using a drilled mirror and being focused into the detection gas of a magnetic bottle electron spectrometer.}
	\label{Fig:setup}
\end{figure}
\begin{figure*}
	\centering
	\includegraphics[width=0.9\textwidth]{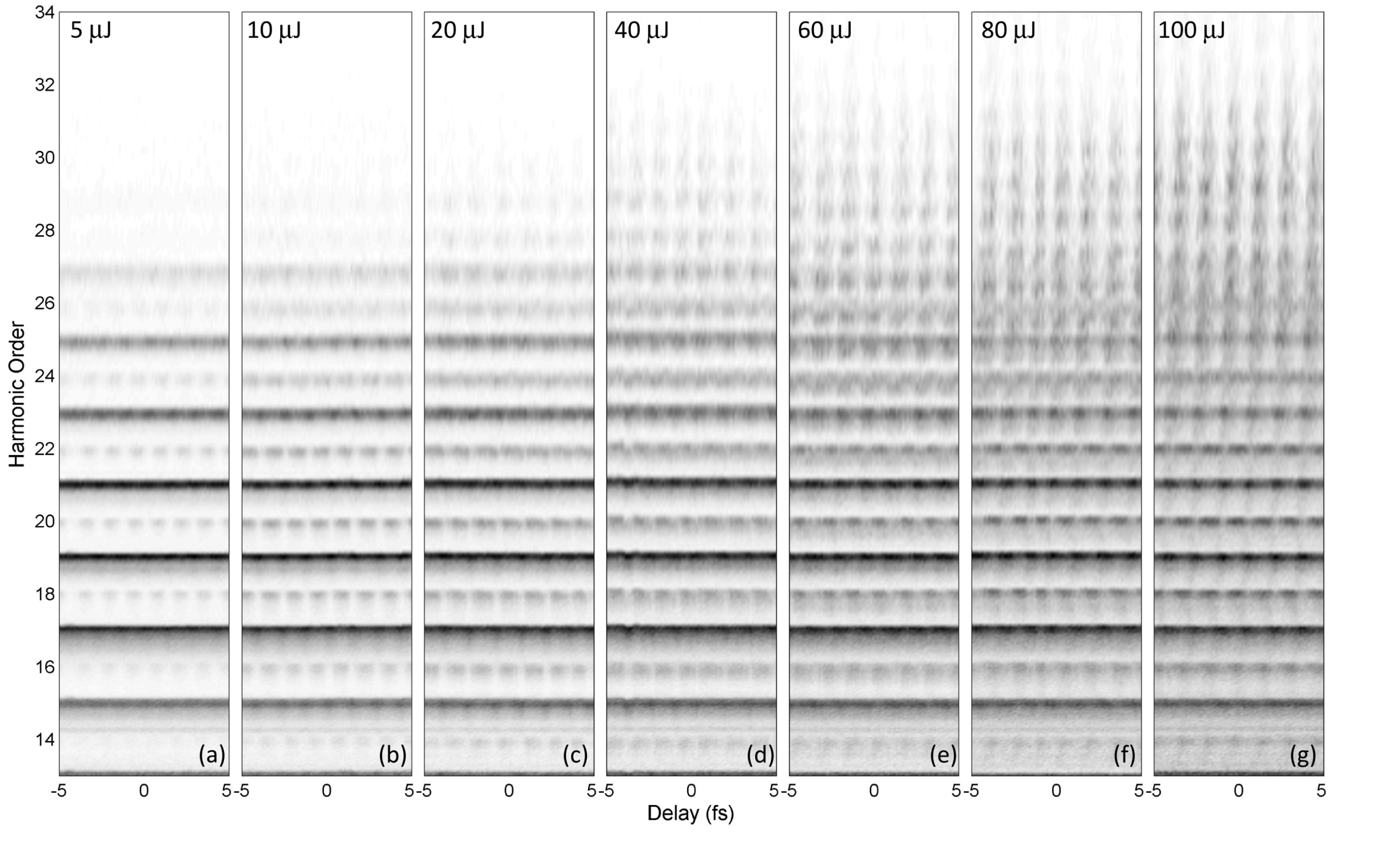}
	\caption{XUV photoionization spectra, recorded for varying sub-cycle delay between IR and XUV field. The corresponding probe pulse energies are written in the upper part of the figures. A change of this energy translates proportionally to a change of the probe intensity. The zero delay indicates the delay at which the maxima of the IR pulse and the APT coincide.}
\label{Fig:Scan}
\end{figure*}
The experiments were performed at the High Power Laser Facility of the Lund Laser Centre. We use a \unit[1]{kHz} chirped-pulse-amplification (CPA) titanium-sapphire based laser system, providing \unit[2]{mJ} \unit[30]{fs} pulses centered around \unit[800]{nm}. The pulses are focused into an Ar gas cell, pulsed at \unit[1]{kHz}, to generate high-order harmonics, which then propagate through silica plates and a thin Al-foil to filter the harmonics to form a well-defined APT (cf. Fig. \ref{Fig:setup}). The APT is then recombined with an IR probe beam in a Mach-Zehnder-type interferometer. In the last step of the interferometer, the APT is passed through a drilled mirror while the IR is reflected on the back side of the mirror. Both beams are focused into the detection gas of a magnetic-bottle electron spectrometer (MBES) with the help of a toroidal mirror. Using a combination of half-wave plate and polarizer in the probe beam, the pulse energy of the IR can be varied continuously between 5 and $\unit[100]{\mu J}$. As all focusing parameters are kept constant, any change in the probe pulse energy will directly result in a proportional change of the IR intensity in the detection region of the MBES. The MBES has a $2\pi$ acceptance angle and a maximum energy resolution of about \unit[100]{meV}.\par
The relative delay of IR probe and APT can be adjusted on two time scales. Using a motorized translation stage in the probe arm we can vary the delay in the range of one femtosecond to several picoseconds. To accurately resolve attosecond processes, a delay stage with a piezoelectric crystal is used to change the relative phase of the two beams with a precision of a few tens of as. This stability allows us to perform interferometry of two-photon pathways as in the RABITT characterization scheme, as shown below. It is also crucial for the use of the AC streak camera method, where the relative phase of the two beams has to be stable with similar precision. \par
\subsection{Delay-Dependent Two-Color Photoionization Spectra}
In the experiments presented here, the resulting photoelectron spectra from an IR probe and XUV field were recorded. A shift in the relative delay of the two fields induces a change in the observed spectrum, and recording the spectra at different subcycle delays allows us, in some conditions, to determine the temporal structure of the XUV emission.\par
\begin{figure}
	\includegraphics[width=0.8\textwidth]{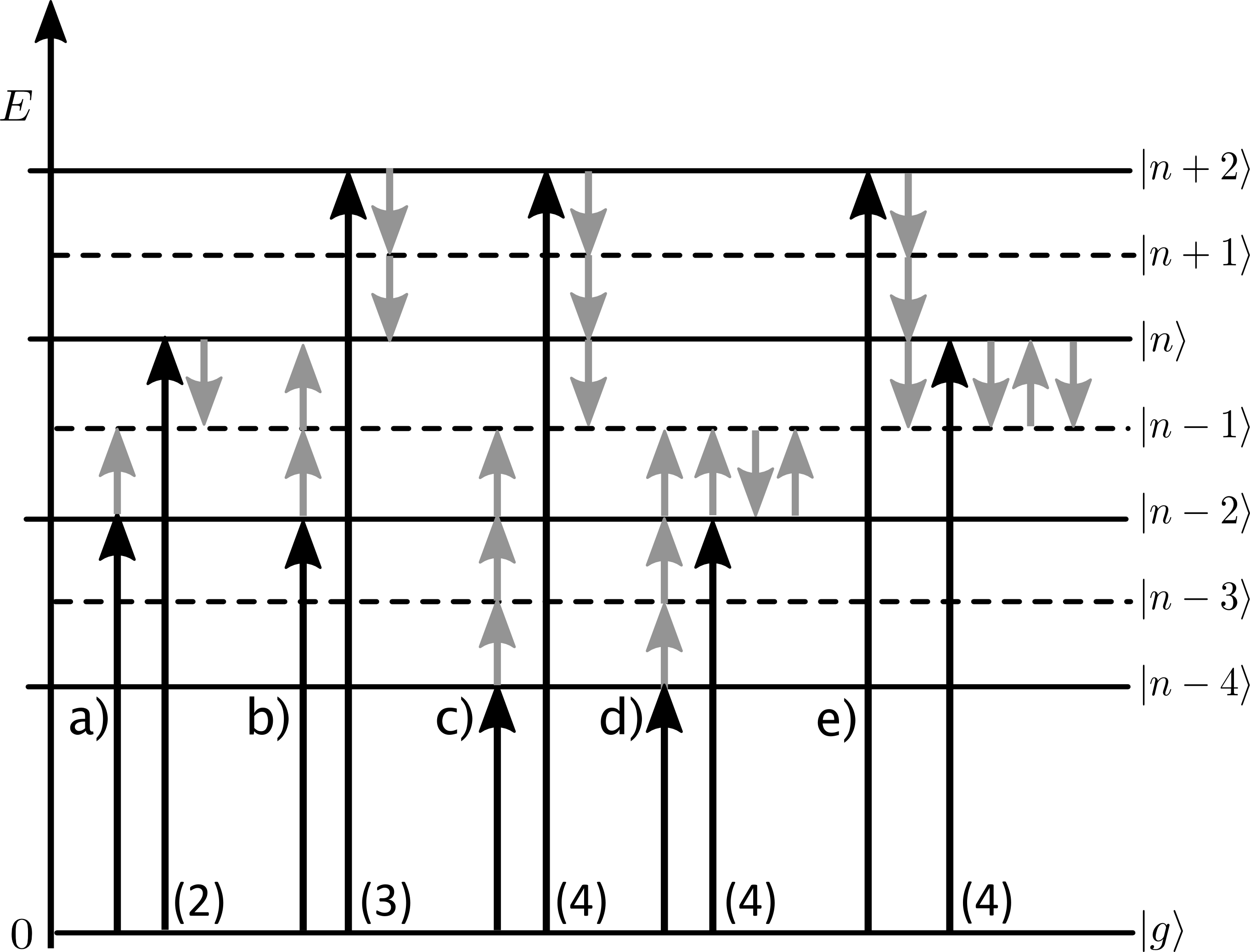}
	\caption{Sketch over possible transitions involving single harmonic- and multiple IR photons. a) shows the RABITT transition, which couples two harmonic photons with an energy difference of $2\omega_R$ to a final state with energy $(n-1)\omega_R$. b) shows that the absorption/emission of two IR photons leads to a similar coupling with odd final energy $n\omega_R$.  c) illustrates that yet another IR photon can be used to couple harmonic photons that are $6\omega_R$ apart in an even final energy $(n-1)\omega_R$. d) and e) shows that three IR photons will also couple new harmonics that are $2\omega_R$ apart in even final energy $(n-1)\omega_R$, which will distort the information from the RABITT signal in a).}
	\label{Fig:PhotonSketch}
\end{figure}

Figure \ref{Fig:Scan} shows the seven scans that constitute our experimental results. From left to right, the probe pulse energy changes by a factor of 20, ranging from 5 to 100\,$\mathrm{\mu J}$. At the lowest intensity, the presence of a weak IR field leads to weak sidebands between the odd harmonic orders. These sidebands originate from $\omega_n+\omega_R$ and $\omega_n-\omega_R$ two-photon transitions \cite{VeniardPRA1996} where $\omega_R$ is the IR photon frequency and $\omega_n = n\omega_R$, with $n$ being an odd integer, see Fig. \ref{Fig:PhotonSketch} a). The two possible pathways to each final state lead to the observed interference pattern. This pattern is repeated every IR half-cycle, $T_{R}/2$, reproducing the frequency of the attosecond pulse periodicity. This probe intensity regime is commonly referred to as the RABITT regime. \par

With increasing IR intensity, the amplitude of the sidebands becomes comparable to direct photoionization by the harmonics [see, e.g. sideband 22 in Figure \ref{Fig:Scan}d)-g)]. These two-photon processes induce a depletion of the peaks at odd harmonic energies, at the delays where the sidebands are maximum. The cutoff region is first affected by depletion since the strength of continuum-continuum transitions rapidly increases with the energy of the initial state [Fig. \ref{Fig:Scan}b) and c) at harmonic 25 and above]. For a probe pulse energy of 40\,$\mathrm{\mu J}$ [Fig. \ref{Fig:Scan}d)], depletion effects become visible even in the low energy region [see harmonic 19 to 23 in Fig. \ref{Fig:Scan}d)-g)].\par
At the higher probe intensities [Figure \ref{Fig:Scan}d)-g)], processes involving more than one IR photon become significant. This is the so-called streaking regime, where the AC-streak camera becomes the preferred characterization method \cite{ItataniPRL2002,KienbergerScience2002,SansoneScience2006}. 
The streaking regime is clearly entered in the last two of the scans in Fig. \ref{Fig:Scan}, the cutoff being increased by as much as seven IR photons (from harmonic 27 and above). Note that the electron signal is still showing discrete IR photon energy spacing because of the periodicity of photoionization from the sequence of attosecond pulses.
\section{Analysis of the Experiment}
\subsection{Reconstruction of Attosecond Beating by Interference in Two-Photon Transitions (RABITT)}
\begin{figure}
	\includegraphics[width=0.65\textwidth]{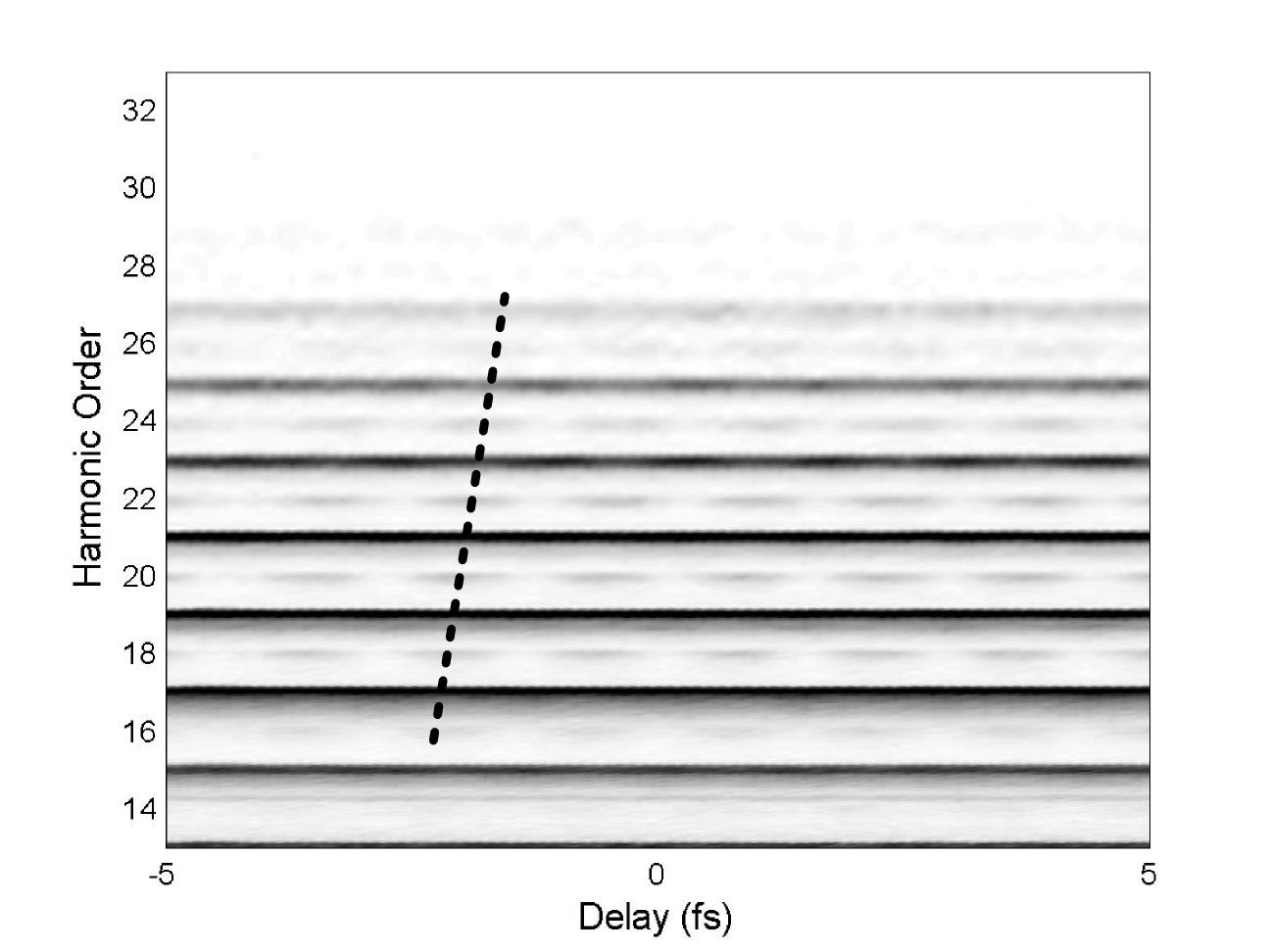}
	\caption{Delay-dependent photoionization spectrum in the RABITT regime, measured at 5\,$\mathrm{\mu J}$ probe pulse energy. The dashed line indicates the sideband maxima.}
	\label{scan1}
\end{figure}

We now analyze the results in Figure \ref{scan1} [same as \ref{Fig:Scan}a)] within a perturbation theory framework. Here, we will consider the photoionization of the detection gas by the APT as a first-order perturbation and the onset of the sidebands and the RABITT modulation as a perturbation of the second order. This is a good description of two-color photoionization at low IR intensities.\par
Contributions to the final state probability amplitudes of order N can be obtained from those of order (N-1) through the following equation \cite{SakuraiMQM1994, BoydNonlinearOptics2003}:
\begin{widetext}
\begin{equation}
a_m^{(N)}(t)=\frac{1}{i\hbar}\sum_f\sum_l\int\limits_0^tdt'\mu_{m,l}E(\omega_f)a_l^{(N-1)}(t')e^{i(\omega_{ml}-\omega_{f})t'},
	\label{eq:1}
\end{equation}
\end{widetext}
where $l$ is the index of the respective contributing states, $\omega_f$ are the frequencies acting on the atom and $m$ denotes the final state. The effect of the IR on the ground state is negligible. The population of continuum states after perturbing the system with a harmonic frequency can be aproximated by 
\begin{equation}
	a_n^{(1)}(t)=\frac{\mu_{n,g}E(\omega_{n})}{i\hbar}t
	\label{eq:a}
\end{equation}
where $\mu_{n,g}$ is the transition dipole matrix element from the ground to the continuum state, and $t$ the interaction time, approximately a measure of the pulse duration. The population of the sideband state $(n-1)$ can be expressed as
\begin{eqnarray}
a_{n-1}^{(2)}(t)&=&\frac{1}{(i\hbar)^2}\frac{t^2}{2}\left[\mu_{n-1,n-2}\mu_{n-2,g}E(\omega_R)E(\omega_{n-2})
+\mu_{n-1,n}\mu_{n,g}E(-\omega_R)E(\omega_{n})\right]\\\nonumber
&\approx&M\frac{E_{R0}E_{X0}}{(i\hbar)^2}\frac{t^2}{2}\left[e^{i(\phi_R+\phi_{{n-2}})}+e^{i(-\phi_R+\phi_{n})}\right]
\end{eqnarray}
where the factor $M$ represents a combination of the various transition dipole moments, which we approximate as equal for both quantum paths, and $E_{R0}$, $E_{X0}$ are the amplitudes of the IR and XUV fields. The sideband intensity will finally oscillate as
\begin{equation}
	I^{(2\omega)}_{n-1}(\varphi)\propto \cos{\left( 2\phi_R+\Delta\phi_{n-2,n}\right)},
	\label{eq:2phi}
\end{equation}
with $\phi_R=\omega_R\tau$, $\tau$ being the delay between the XUV and the IR fields. $\Delta\phi_{n-2,n}$ is the phase difference between the harmonic spectral phases $\phi_{{n-2}}$ and $\phi_{{n}}$. This allows us to obtain the phase difference $\Delta\phi_{n-2,n}$ from a Fourier transform of the spectrum over delay.\par
\begin{figure}
	\includegraphics[width=0.40\textwidth]{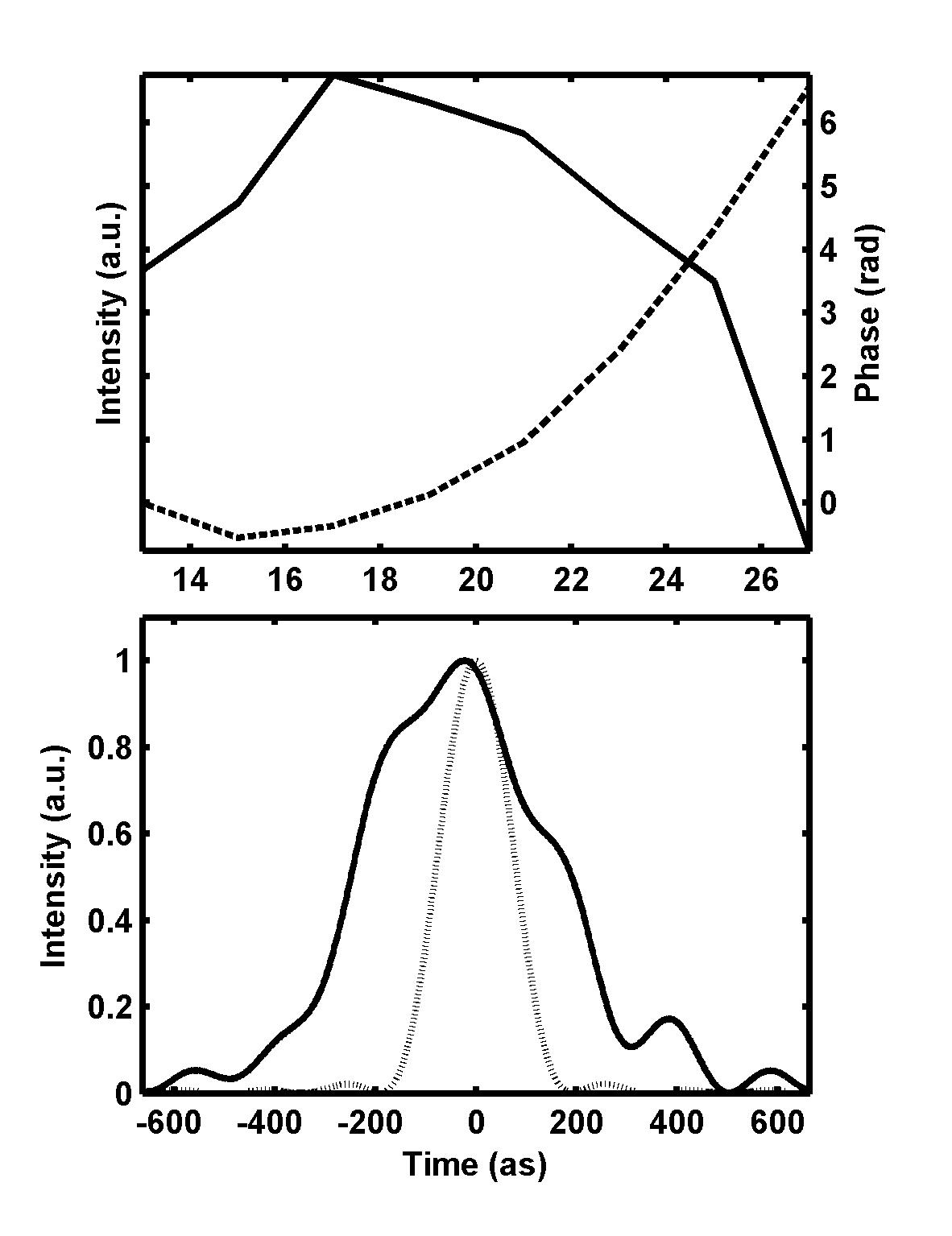}
	\caption{Reconstruction of average pulse in the pulse train, measured by the RABITT method. The upper panel shows the harmonic intensities (solid line) and integrated spectral phase (dashed line). Using these intensities and phase, one can reconstruct the pulse shape as in the lower panel. The Fourier limited pulse shape is shown as dashed.}
	\label{Fig:rec5rab}
\end{figure}

Using the obtained phase, we can reconstruct a pulse shape as in Figure \ref{Fig:rec5rab}. The Fourier limit for our spectrum is 160\,as. Due to the chirp rate of 18400\,$\mathrm{as^2}$ we obtain an average pulse duration of 440\,as FWHM. This method of measuring APTs has been successfully employed in a number of experiments \cite{PaulScience2001,MairesseScience2003,LopezMartensPRL2005}. The high chirp rate and asymmetric pulses come as no surprise as only one thin aluminum filter was employed to counter the intrinsic chirp of the HHG process. The addition of more filters would help to approach the Fourier limit and further compress the pulses \cite{LopezMartensPRL2005,VarjuLP2005}.
The RABITT method provides access to the relative phase difference of presumably monochromatic harmonics \cite{VeniardPRA1996,PaulScience2001}. The measured phase difference is equal to the group delay (GD) over the spectrum of the APT, and an integration allows us to reconstruct an average attosecond pulse in the train. Such a retrieved pulse represents a good approximation for pulses within the FWHM of the APT, accounting for about 90\% of the signal from experiments with such trains. There are extensions and implementations of the method that yield more information on the full structure of the APT \cite{VarjuPRL2005,MauritssonJPB2005}.\par
\subsection{Intensity Dependence of RABITT Signal}
\begin{figure}
	\includegraphics[width=0.7\textwidth]{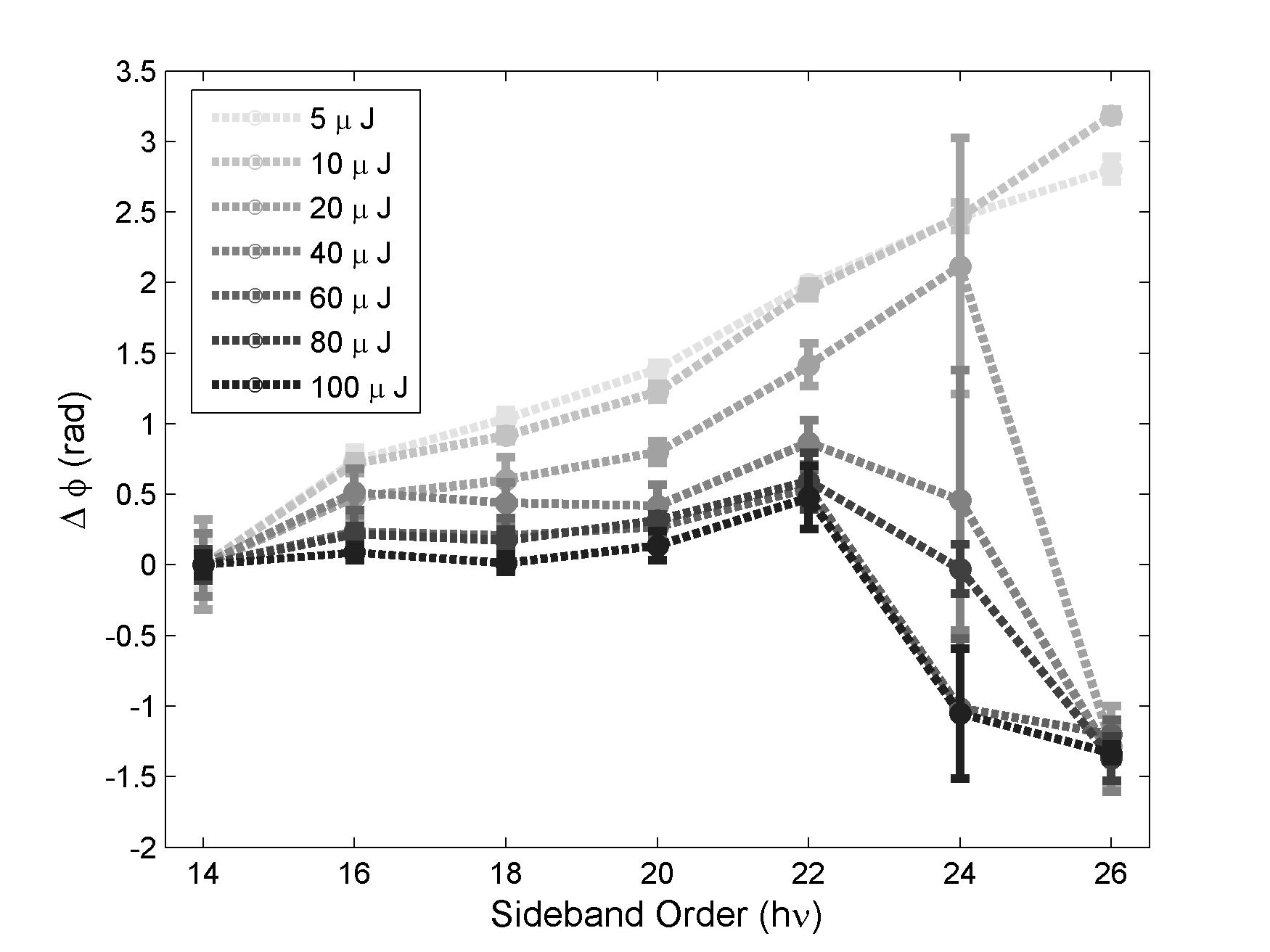}
	\caption{Phase differences $\Delta \phi$ measured over the harmonic spectrum, for different probe pulse energies. Higher probe intensities result in a flatter measured spectral phase in the plateau region (order 14-22).}
\label{Fig:Rabitt}
\end{figure}
The influence of an intense probe field on the measured spectral phase was studied by progressively increasing the IR intensity and recording the delay-dependent photoionization spectra. Figure \ref{Fig:Rabitt} shows the phase differences $\Delta\phi_{n-2,n}$ evaluated according to the method outlined above, for varying probe intensities. It is clear that the slope is decreasing with increasing intensity, as if the chirp rate was lower, even though the actual APT remained the same.\par
 As a consequence of the phase variation with intensity shown in Figure~\ref{Fig:Rabitt}, the reconstructed pulses will appear more compressed at higher probe intensities, which is a pure artifact of the increased IR intensity.
The RABITT analysis relies heavily on the presence of a single IR photon contributing to the final state as to allow only the phase difference of the neighboring harmonics to be measured. Higher-order processes perturb the RABITT signal, as will be discussed in more detail in Section \ref{sec:fourier}.\par
\subsection{Streaking}
\begin{figure}
	\includegraphics[width=0.65\textwidth]{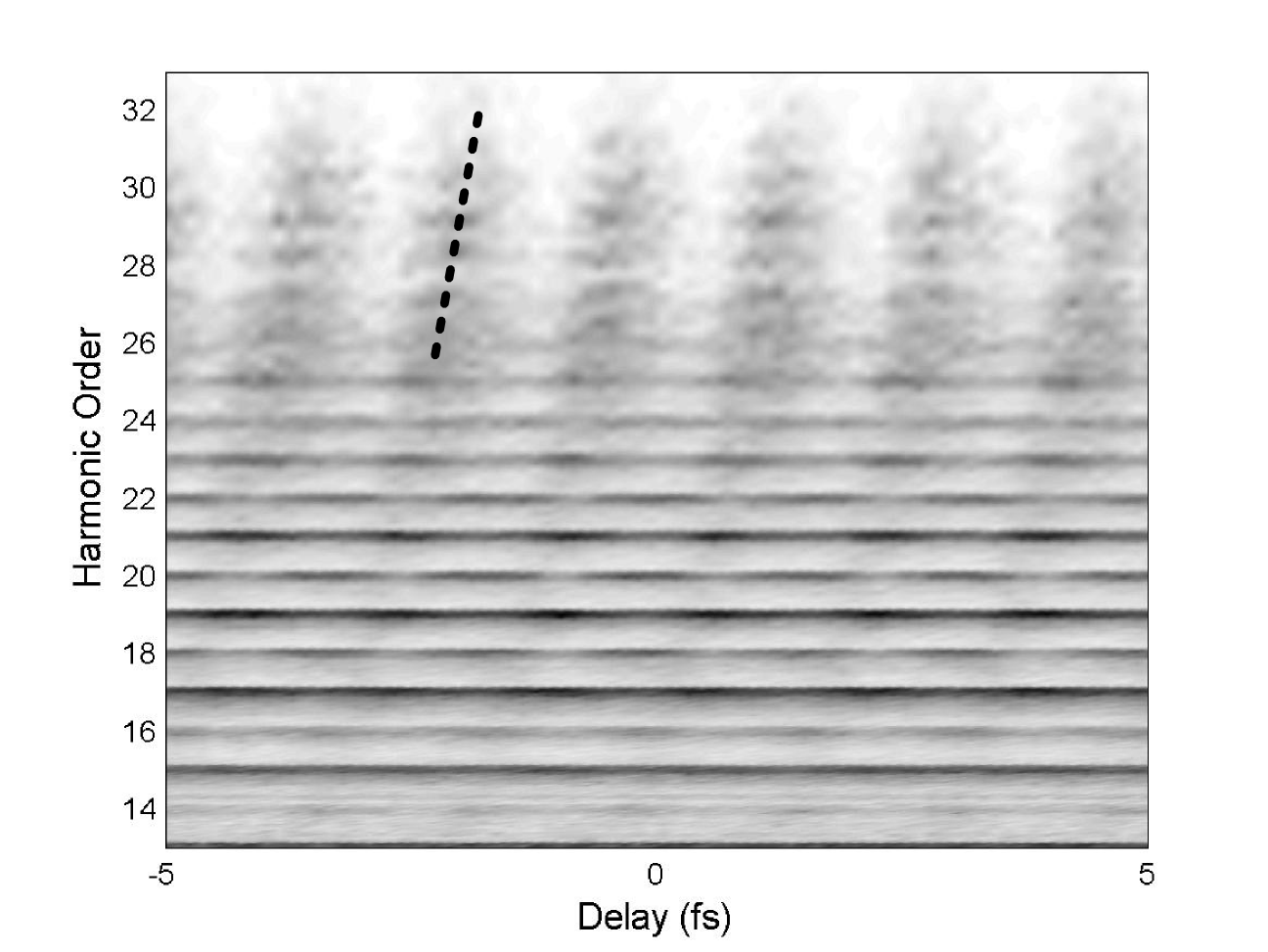}
	\caption{Delay-dependent photoionization spectrum in the streaking regime, measured at 100\,$\mathrm{\mu J}$ probe pulse energy. }
	\label{scan7}
\end{figure}

We now consider briefly the regime of high probing intensities [Figure \ref{scan7}, same as \ref{Fig:Scan} g)] where the interaction between photoelectrons and probing laser field can be understood by simple classical arguments. 
In this regime, an electron is released into the continuum at a given ionization time $t_i$ due to photoionization by an attosecond pulse. It will gain an additional momentum proportional to the IR vector potential $\vec{A}(t_i)$ at the time of ionization. This momentum shift imparted by the IR probe is therefore dependent on the relative delay between the two fields. With APTs generated by two-color fields, or with single attosecond pulses, the pulse properties (duration and chirp) can be extracted from analyzing the streaking trace and the pulse(s) can be reconstructed \cite{ItataniPRL2002,KienbergerScience2002,SansoneScience2006}. In the case of an APT with two pulses per cycle, the situation is more complex and a reconstruction of the attosecond pulses from the experimental data is more difficult, since the sign of the probe field reverses for consecutive pulses. Without going further with the analysis of our experimental data, we note that the chirp of our attosecond pulses, however, is clearly visible in the high energy region of the spectrum [see dashed line in Figure \ref{scan7}].
The onset of the streaking regime also explains the observed phase shift of the sideband orders close to the cutoff in Figure \ref{Fig:Rabitt}. In the perturbative regime and, for simplicity, in absence of attosecond chirp, the maxima of the sidebands occur at the maxima of the probe laser electric field. At moderate probe intensity, depletion of the harmonic states will lead to maxima of the harmonics shifted by $\pi/2$ with respect to the sidebands, thus corresponding to the zeros of the electric field. At high energy, electron peaks, coming from direct electrons being streaked by the probe field, will therefore be maxima at the zeros of the electric field, thus shifted by a factor $\pi/2$ from the sidebands. 
\section{Fourier Decomposition of the Photoelectron Spectra\label{sec:fourier}}
\subsection{Perturbation Theory Analyzis}
To better understand the influence of a high IR intensity on a RABITT measurement, we extend the analyzis performed in the previous section, based on perturbation theory to the next orders. The third order of perturbation includes components coupling the states $(n+2)$ and $(n-2)$ to the harmonic $(n)$ state by two IR photon absorption or emission [Figure \ref{Fig:PhotonSketch}b)]:
\begin{widetext}
\begin{eqnarray}
a_{n}^{(3)}&=&\frac{1}{(i\hbar)^3}\frac{t^3}{6}\left[\mu_{n,n-1}\mu_{n-1,n-2}\mu_{n-2,g}E(\omega_R)E(\omega_R)E(\omega_{n-2})+\mu_{n,n-1}\mu_{n-1,n}\mu_{n,g}E(\omega_R)E(-\omega_R)E(\omega_{n})\right.\nonumber\\\nonumber &+&
\left.\mu_{n,n+1}\mu_{n+1,n}\mu_{n,g}E(-\omega_R)E(\omega_R)E(\omega_{n})+\mu_{n,n+1}\mu_{n+1,n+2}\mu_{n+2,g}E(-\omega_R)E(-\omega_R)E(\omega_{n+2})\right]\nonumber\\
&\approx&M\frac{E_{R0}^2E_{X0}}{(i\hbar)^3}\frac{t^3}{6}\left[e^{i(2\phi_R+\phi_{n-2})}+2e^{i\phi_{{n}}}+e^{i(-2\phi_R+\phi_{{n+2}})}\right].
	\label{eq:2}
	\end{eqnarray}
\end{widetext}
This term gives rise to oscillations at frequencies $4\omega_R$, $2\omega_R$, and to a constant term. The intensity of the $4\omega_R$ modulation at the harmonic energy $n\omega_R$ is
\begin{equation}
	I_{n}^{(4\omega)}(\phi_R)\propto\cos{(4\phi_R+\Delta\phi_{n-2,n+2})}
	\label{eq:4phi}
\end{equation}
which is similar to Equation \ref{eq:2phi}. The next order of perturbation leads to a $6\omega_R$ component in the sideband states and the next to an $8\omega_R$ component in the harmonic states. Higher IR intensities lead to new couplings of states lying further apart and thus higher modulation frequencies.\par
Figure \ref{Fig:PhotonSketch} illustrates how the number of pathways that lead to a given modulation frequency is unique only under certain restrictions. Even numbers of contributing IR photons couple to harmonic states while odd numbers couple to sideband states. The $4\omega_R$ frequency is meaningful only in the harmonic final states, as shown in Eq. \ref{eq:2} and the phase difference that can be determined is that between harmonic states $(n-2)$ and $(n+2)$. A Fourier transform that isolates this component will allow us to access to the spectral phase in the same way as with the conventional RABITT, looking at the $2\omega_R$ modulation in the sideband states. Experimentally, due to the resolution of our scans, the $6\omega_R$ component is the highest resolvable. It is already close to the Nyqvist limit, with only three to four data points per period.\par
\begin{figure}
	\includegraphics[width=0.65\textwidth]{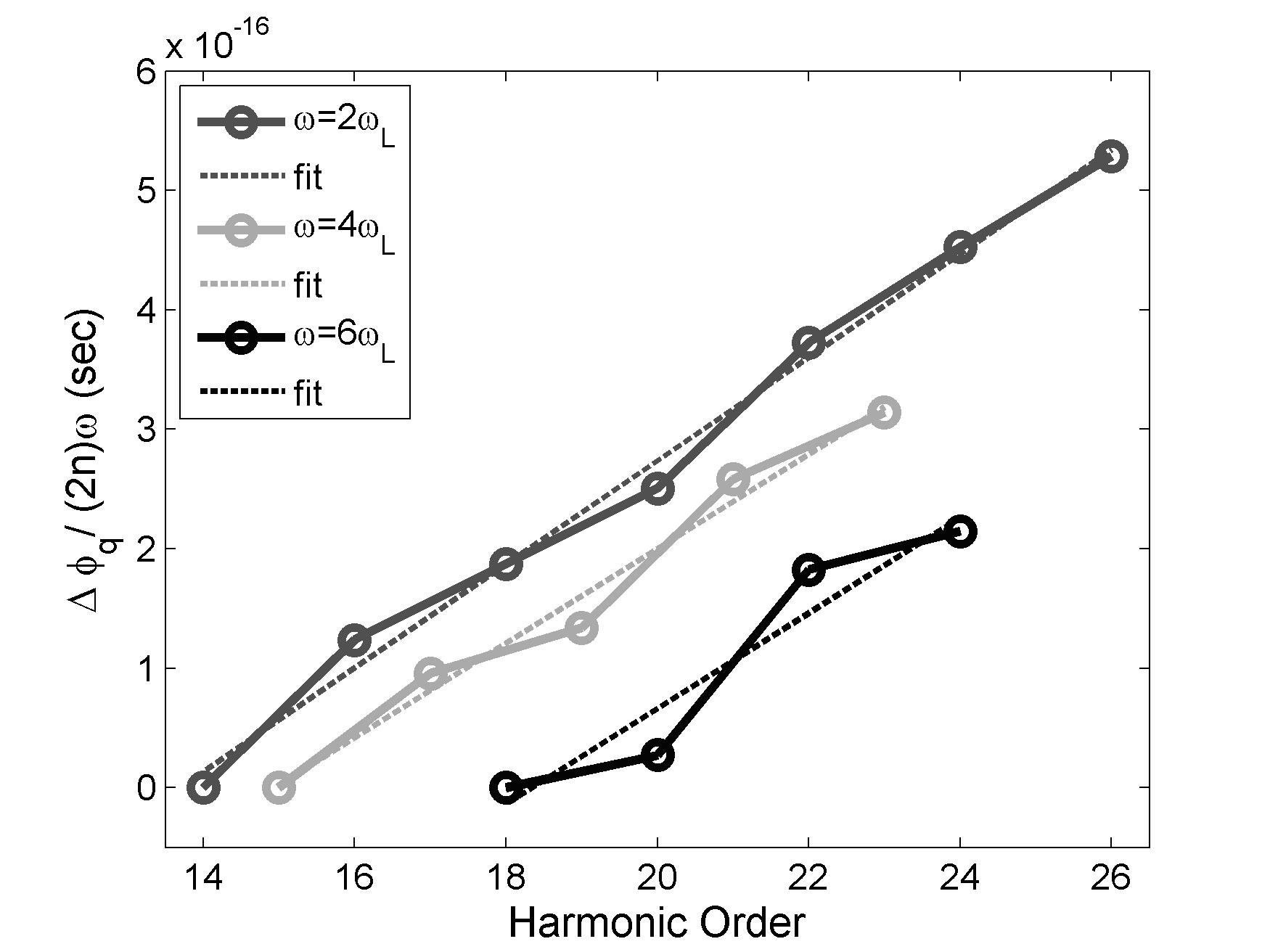}
	\caption{Comparaison of the obtained phase differences for three different frequency components present in the experimental electron signal. The conventional RABITT includes contribution from sidebands 14 to 26. The $4\omega_R$-component has been extracted from harmonics 15 to 23 and the $6\omega_R$-modulation was obtained from sidebands 18 to 24. The curves have been shifted for better comparison.}
	\label{Fig:rabcomp}
\end{figure}
Fig. \ref{Fig:rabcomp} shows a comparison of the group delay obtained over the spectrum for three different modulation frequencies. The usual RABITT is based on the 5\,$\mathrm{\mu J}$-scan, the phases of the $4\omega_R$-modulation were obtained at 20\,$\mathrm{\mu J}$, and the 6$\omega_R$-modulation was present for the central harmonics in the 80\,$\mathrm{\mu J}$-scan.  Excellent agreement was found between the chirp rates obtained with the different methods. Our results are also consistent with previous experiments \cite{MairesseScience2003,LopezMartensPRL2005}. From the $2\omega_R$-measurement we obtain a chirp rate of 18400\,$\mathrm{as^2}$ or - in terms of a group delay - an emission time difference $\Delta t_e=\Delta\phi/(2\omega_R)=87$\,$\mathrm{as}$. This is very similar to 16800\,$\mathrm{as^2}$ ($\Delta t_e=79$\,$\mathrm{as}$) obtained from the $4\omega_R$-component in the harmonics. The slope of the phase difference in the $6\omega_R$-component is also in good agreement with 16900\,$\mathrm{as^2}$ chirp rate or a $\Delta t_e$ of 80\,$\mathrm{as}$. \par
\subsection{Generalization}
Because of the analogy between the conventional RABITT and the possibility to extract the same information from higher components we will now introduce a more general expression for the delay dependence of our photoelectron spectrum. This expression will hold beyond the breakdown of the perturbation picture. The amplitude of any photoelectron peak in the spectrum is periodic with delay $\phi_R$ between the IR and XUV fields. It can therefore be expressed as a discrete Fourier series:
\begin{equation}
	S(\phi_R,I_R)=\sum_{k=-\infty}^{\infty}\tilde{S}_{k}(I_R)e^{i\phi_R 2k}=\sum_{k=0}^{\infty}{S}_{k}(\phi_R,I_R)
	\label{eq:3}
\end{equation}
with $\tilde{S}_k^*=\tilde{S}_{-k}$. Considering only the $k=\pm 1$ contributions we have
\begin{eqnarray}
S_1(\phi_R,I_R)&=&\tilde{S}_1(I_R)e^{i\phi}+\tilde{S}_1^*(I_R)e^{-i\phi}\\\nonumber
&=&2\left|\tilde{S}_1(I_R)\right|\cos\left(2\phi_R+arg(\tilde{S}_1(I_R))\right)
	\label{eq:4}
\end{eqnarray}
which is the common RABITT case, for low IR intensity, with $arg(\tilde{S}_1(I_R))=\Delta \phi_{n-1}$. For higher IR intensities, higher order terms contribute, the argument will change and higher order terms of the sum $\tilde{S}_{k}(I_R)$ will grow revealing higher contributing frequencies. In the case of $k=2$,
\begin{equation}
	S_2(\phi_R,I_R)=2\left|\tilde{S}_2(I_R)\right|\cos\left(4\phi_R+arg(\tilde{S}_2(I_R))\right),
	\label{eq:5}
\end{equation}
which can readily be identified as the $4\omega_R$ component with an argument of $\Delta \phi_{n-2,n+2}$ for a suitable probe intensity. Figures \ref{Fig:PhotonSketch}d)-e) show the effect of the participation of three IR photons, and how two non-surrounding harmonics contribute their phase difference to sideband $n-1$, destroying the unique source of the $2\omega_R$ modulation. Thus $arg(\tilde{S}_1(I_R))$ changes and the phase observed in the RABITT analysis varies with IR intensity.\par

At the same order of perturbation, a $6\omega_R$ modulation can be constructed with three IR photons  coupling harmonic $n-4$ and $n+2$ to sideband $n-1$ [cf. Fig. \ref{Fig:PhotonSketch} d)]. For sufficiently high intensity, the number of contributing IR photons is $\gg 1$, resulting in a great number of frequencies added to form the delay-dependent signal in the sideband. More and more harmonics contribute to the phase, which becomes flat. The oscillation does not uniquely depend on the adjacent orders anymore - the requirement for RABITT. This is the streaking regime, where the classical limit is reached.\par
\begin{figure}
	\includegraphics[width=0.65\textwidth]{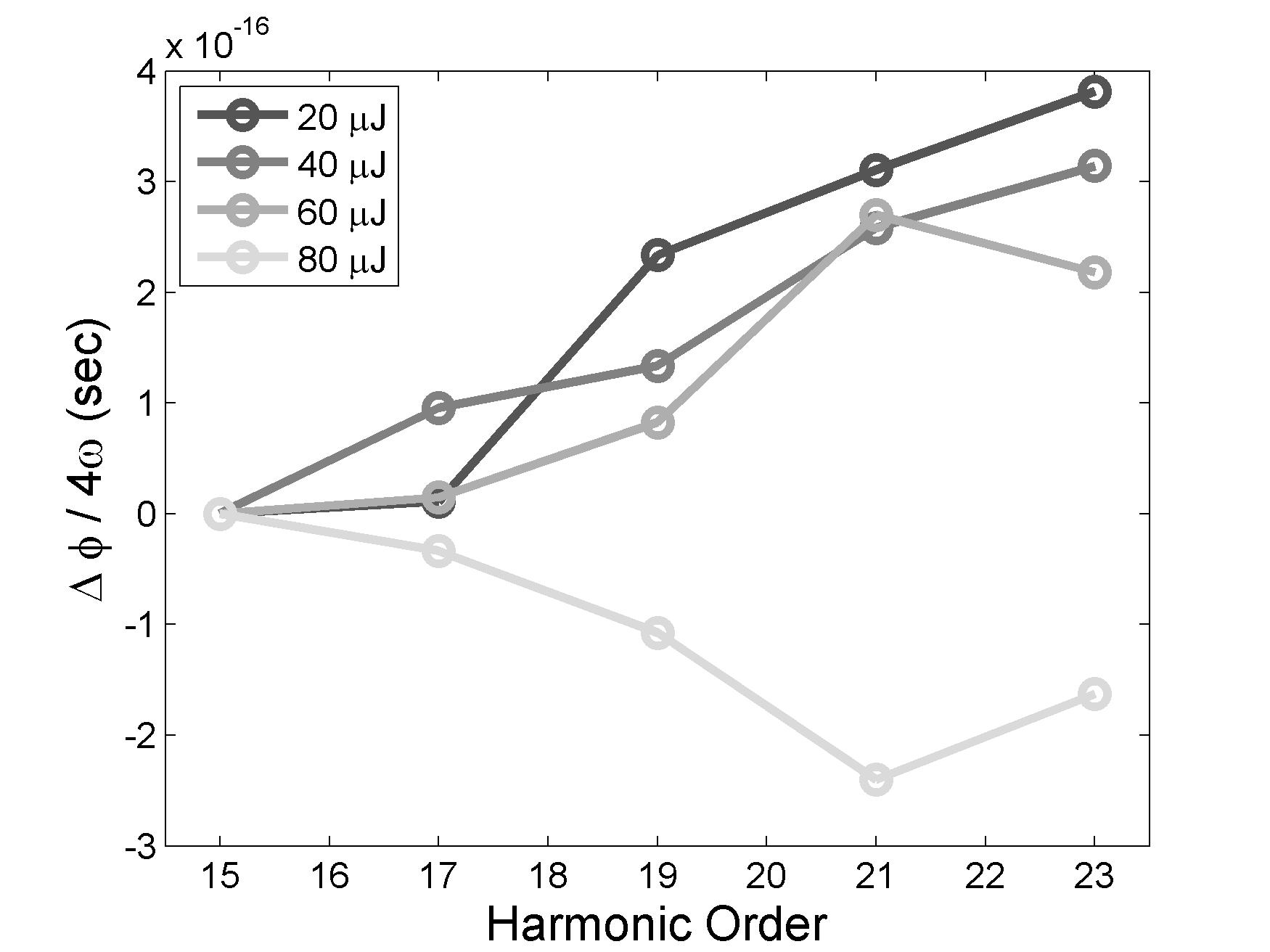}
		\caption{Intensity dependence of the measured phase from the $4\omega_R$ component in the harmonics. The measured chirp rate over the spectrum shows good agreement at 40\,$\mathrm{\mu J}$ with the $2\omega_R$-measurement from the sidebands. It breaks down at higher probe energy.}
\label{Fig:2hrab}
\end{figure}

As we previously studied the effect of the IR intensity on the phase measurement in the conventional RABITT method, we perform a similar analysis for our measurements with the $4\omega_R$ component. Figure \ref{Fig:2hrab} shows the intensity dependence of the harmonic phase measured in this component. We find that at 20 to 60\,$\mathrm{\mu J}$ the phase is meaningful compared with the conventional measurement. The measurement at 80\,$\mathrm{\mu J}$ is showing a similar characteristic flattening of the phase as the conventional RABITT. The $4\omega_R$ breakdown occurs for much higher probe intensities. In this case the number of contributing photons becomes so high that $arg(\tilde{S}_2(I_R))$ is affected by ambiguous coupling of states. The analyzis of this $4\omega_R$ component provides an additional check on the measured phase from conventional RABITT method. This allows us to assess whether the probe intensities in any experiment have been too high by comparing the $2\omega_R$ with the $4\omega_R$ components and possibly even higher orders.\par
\section{Conclusion}
We have studied the influence of IR probe intensity in two-color XUV photoionization experiments. In the laser-assisted ionization process that is at the heart of the RABITT method for characterizing APTs, the probe intensity needs to be maintained at levels which only perturb the process very slightly. A high probe intensity will greatly alter the measured phase relation of the individual high-order harmonics, making it appear flat over the spectrum, as if the pulses were compressed. RABITT is only valid in the limit of weak IR intensity. \par
In a next step, the increasing number of contributing IR photons was tracked down by Fourier-analysis of sideband and harmonic states. As an increasing number of photons allows to couple states further and further away, higher modulation frequencies occur. The $4\omega_R$ component of the harmonic states allows us to obtain their mutual phase relation. Also in this case an increasing IR intensity started to affect the phase measurement making the pulses appear artificially compressed. We believe these experiments will allow the scientific community to gain a better understanding of the IR intensity dependence of laser-assisted ionization of a gas using high-order harmonics.\par
This paper is dedicated to the memory of N. B. Delone, who was a father of this exciting field of research ``atoms in strong laser field'', with the first ``multiphoton ionization'' (1965) and ``multielectron multiphoton ionization'' (1975) experimental results. This research was supported by the Marie Curie Intra-European Fellowship (Attotech), the Marie Curie Early Stage Training Site (MAXLAS), the Knut and Alice Wallenberg Foundation, the Swedish Research Council, and the National Science Foundation (Grant No. PHY-0701372). 

\hyphenation{Post-Script Sprin-ger}

\end{document}